\documentclass[twocolumn,showpacs,amsmath,amssymb,prl]{revtex4}

\usepackage{graphicx}
\usepackage{dcolumn}
\usepackage{bm}

\newcommand{\ket}[1]{\vert#1\rangle}

\begin{document}

\title{Deterministic unitary operations on time-bin qudits for quantum communication}

\author{F\'elix Bussi\`eres}
\altaffiliation[Also at ]{Laboratoire d'informatique th\'eorique et quantique,
Universit\'e de Montr\'eal, C.P. 6128, Succ. Centre-Ville, Montr\'eal (QC), H3C~3J7 Canada}
\email{felix.bussieres@polymtl.ca}
\author{Yasaman Soudagar}
\author{Guido Berlin}
\author{Suzanne Lacroix}
\author{Nicolas Godbout}
\affiliation{Centre d'optique, photonique et lasers, Laboratoire des fibres optiques, D\'epartement de g\'enie physique, \'Ecole Polytechnique de Montr\'eal, C.P. 6079, Succ. Centre-ville, Montr\'eal (QC), H3C~3A7 Canada}

\date{\today}

\begin{abstract}
We show for the first time how deterministic unitary operations on time-bin qubits encoded in single photon pulses can be realized using fiber optics components that are available with current technology. We also generalize this result to operations on time-bin qudits, i.e. $d$-level systems, and show that this can be done efficiently using 2$\times$2 beamsplitters and phase modulators. Important benefits for experimental quantum communication are highlighted. This work shows how to bridge the gap between current proof-of-principle demonstrations and complete, deterministic experiments.
\end{abstract}


\maketitle

Quantum information science is rooted in the idea that information can be encoded and processed in qubits, i.e.~two-level quantum systems~\cite{NC00}. Recently, this fruitful proposition led to important theoretical results in the field of quantum communication along with experimental demonstrations of quantum cryptography~\cite{BB84}, quantum repeaters~\cite{BDCZ98}, tests of Bell's theorem~\cite{Bell64} and quantum teleportation~\cite{BBCJPW93} (see Ref.~\onlinecite{TW01} for a review). Moreover, there are now numerous proposals benefiting from the use of qudits ($d$-level quantum systems). In cryptography, qudits increase the tolerance to noise~\cite{CBKG02} and in quantum communication, they improve channel capacity~\cite{FTMS03}. When entangled, they enhance the security bounds for coin flipping~\cite{Ambainis01}, they reduce the communication complexity below the classical limit~\cite{BZZ02} and they give a solution to the Byzantine agreement problem~\cite{FGM01}. Entangled qudits are also useful for testing Bell's theorem by providing a stronger violation~\cite{KGZMZ00} and a greater robustness against noise~\cite{CGLMP02}.

\begin{figure}[!h]
\includegraphics[scale=1]{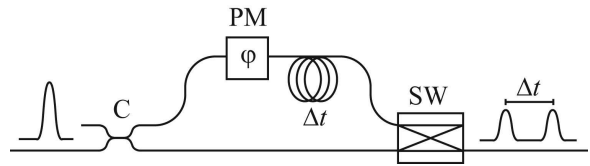}
\caption{\label{fig:time-bin} Experimental set-up to generate time-bin qubits. Lines are optical fibers, C is a coupler (the all-fiber equivalent of a beamsplitter), PM is a phase modulator and SW is a 2$\times$1 optical switch.}
\end{figure}

Photons and optical fiber are natural candidates to implement quantum communication protocols and for this purpose, the time-bin encoding for qubits is proven to be very robust and practical~\cite{B92,BGTZ99}. To generate a time-bin qubit, a single photon pulse is first split in two rails using a 2$\times$2 coupler (the all-fiber equivalent of a beamsplitter) as shown on Fig.~\ref{fig:time-bin}. Then, each component tra\-vels through either the ``short'' or ``long'' rails. Finally, both are time-multiplexed in the same fiber using a 2$\times$1 optical switch. At the output, the single photon is in a coherent superposition of being in two time-bins separated by a delay $\Delta t$: $\ket{\psi} = \alpha \ket{\text{short}} + \beta\ket{\text{long}}$. By changing the coupling ratio of the coupler and the phase delay $\varphi$ of the phase modulator, all pure qubit states can be ge\-ne\-ra\-ted. Time-bin entanglement in the state $\alpha \ket{\text{short, short}} + \beta \ket{\text{long, long}}$ can also be generated efficiently using spontaneous parametric downconversion~\cite{BGTZ99}. This type of entanglement has been shown to be robust against po\-la\-ri\-za\-tion mode dispersion, phase fluctuations and chromatic dispersion over at least 50~km of transmission in optical fiber~\cite{MRTZLG04,FGRZ04}.

The main difficulty with time-bin encoding is that both single qubit unitary operations and measurements in any basis are not trivial to implement. Actually, most of the experiments done so far use non-deterministic operations with post-selection~\cite{GRTZ02,MRTZG03,TW01} and this reduces the success rate, if not leading to a complete failure. For example, complete quantum teleportation of a time-bin qubit cannot be achieved wi\-thout using deterministic single qubit operations to correct the teleported state~\cite{MRTZG03}. A similar problem exists in the experiments testing Bell's theorem with time-bin qubits as we discuss below. Also, multi-user protocols requiring cascaded operations on the same qubit are not scalable when limited to probabilistic operations. The ubiquitous need for single time-bin qubit and qudit deterministic operations for quantum communication and the lack of any general scheme to do this are the motivations of this work. In this letter, we extend on our work of Ref.~\onlinecite{BSBLG06:proceedings} and propose different experimental schemes to perform de\-ter\-mi\-nis\-tic unitary operations on time-bin qubits and qudits using photons. These schemes can be implemented with technology that is currently available. Throughout the text, the schemes are presented with single mode optical fiber circuits and wavelengths around 1550~nm. Nevertheless, they can be directly adapted to any kind of waveguide or free-space transmission.

All unitary operations on a single polarization qubit are easy to implement using an all-fiber polarization controller. The first scheme we propose takes advantage of this by conver\-ting a time-bin qubit to a polarization qubit to perform the operation and then converting back to a time-bin qubit for transmission.
\begin{figure}[!h]
\includegraphics[scale=1]{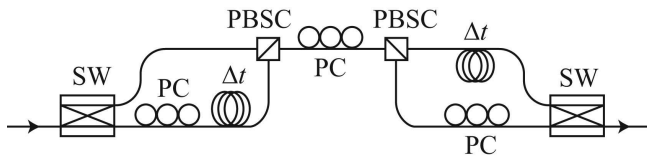}
\caption{\label{fig:polarization-gate} Experimental layout of the time-bin gate using polarization. Lines are optical fibers, SW is a 1$\times$2 optical switch, PC is a polarization controller and PBSC is a polarization beam splitter/combiner.}
\end{figure}
The gate is shown on~Fig.~\ref{fig:polarization-gate}. First, a single photon in a time-bin qubit state with a given polarization, say horizontal (H), is incident from the left on the 1$\times$2 switch which routes the $\ket{\text{short}}$ and the $\ket{\text{long}}$ components to the lower and upper rails res\-pec\-ti\-vely. The polarization of the lower rail is flipped to vertical (V) by a polarization controller and is also delayed by $\Delta t$. This delay synchronizes the $\ket{\text{short}}$ and $\ket{\text{long}}$ components at the inputs of the polarizing beam splitter/combiner (PBSC). The latter reflects and transmits V and H polarizations, hence, the time-bin qubit is converted to a single-rail polarization qubit according to the mapping $\ket{\text{short}} \rightarrow \ket{V}$ and $\ket{\text{long}} \rightarrow \ket{H}$. Then, the output of the PBSC is fed into an all-fiber polarization controller that transforms any polarization to any other with negligible loss. After the rotation, the polarization qubit is converted back to an H~polarized time-bin qubit using the inverse of the operation done in the first section. Therefore, this gate can implement all unitary operations on a single time-bin qubit.

To implement a given operation, the switches are the only active components. The faster the switching speed, the smaller is the required separation between the $\ket{\text{short}}$ and $\ket{\text{long}}$ components, thus, the smaller is the required path length difference. With current technology, electrooptic material allow for switching frequencies of at least 10~GHz, hence, the path length difference can be set at about 2~cm in standard single mode fiber. This is crucial since the relative phase between the branches has to be stabilized in temperature. A path difference of 2~cm would require the temperature to be stable within 1~tenth of a degree Kelvin at room temperature, which is not difficult to achieve in the laboratory. The two PBSC can be fabricated with fused fiber couplers and are therefore practically lossless~\cite{Diet05}. The major limiting factor is the insertion loss of the switches caused mainly by the mode mismatch between the fiber core and the waveguide of the switch. With current technology, this loss can be lowered down to about 1.5~dB for a total of 3~dB for the whole gate. 

\begin{figure}[!h]
\includegraphics[scale=1]{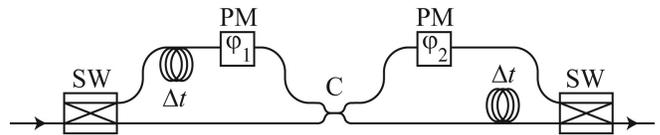}
\caption{\label{fig:dual-rail-gate} Experimental layout of the time-bin gate using dual-rail encoding. Lines are optical fibers, SW is a 1$\times$2 optical switch, PM are phase modulators and C is a coupler.}
\end{figure}

The second scheme for deterministic operations we propose works by converting the time-bin qubit to a dual-rail qubit for processing~\cite{CY95}. The gate is pictured in~Fig.~\ref{fig:dual-rail-gate} and works as follows. A time-bin qubit is incident from the left on a 1$\times$2 optical switch that routes the $\ket{\text{short}}$ and the $\ket{\text{long}}$ components to the upper and lower rails res\-pec\-ti\-vely. 
Next, the upper rail is delayed by $\Delta t$ to synchronize $\ket{\text{short}}$ with $\ket{\text{long}}$ and consequently, the photon is converted to the dual-rail encoding: $\ket{\text{short}} \rightarrow \ket{1}_u\ket{0}_l$ and $\ket{\text{long}} \rightarrow \ket{0}_u\ket{1}_l$, where the labels $u$ and $l$ stand for the upper and lower rails respectively. The two components of the single photon interfere at the coupler set to a specific coupling ratio to implement a given gate. This, supplemented with two phase modulators, can implement any unitary operation on the dual-rail qubit. The remaining part of the gate converts the dual-rail qubit back to a single-rail time-bin qubit. Therefore, this gate can implement all unitary operations on a single time-bin qubit.

To implement a given operation, the switches are the only active components and the temperature requirements are the same as in the previous scheme. Moveover, stable and precise phase modulators can be made lossless using piezoactuators glued on the fiber. Finally, the typical loss of an all-fiber coupler is of the order of 0.1~dB or less. Therefore, the only important sources of loss are the switches and again, it can be lowered down to about 3~dB in total. The main advantage of this scheme over the previous one is that it can be made insensitive to alignment if pola\-ri\-zation independent switches are used. Also, this scheme is well suited for tasks that require ultra-fast and deterministic feedforward processing like quantum teleportation. Indeed, an ultra-fast tunable coupler can be made with a Mach-Zehnder interferometer using an eletrooptic phase modulator in one of its arms, hence yielding a single circuit that can implement all operations. Using theses ideas, we have shown how the cluster state model of quantum computation can be implemented in optical fiber circuits~\cite{SBBLFG06-paper}.

\begin{figure}[!h]
\includegraphics[scale=1]{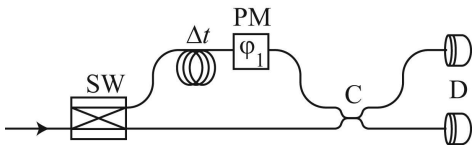}
\caption{\label{fig:measurement} Experimental layout of the time-bin gate to measure in any desired basis. Lines are optical fibers, SW is a 1$\times$2 optical switch, PM is a phase modulator, C is a coupler and D is a detector.}
\end{figure}

We now discuss how this scheme is useful for testing Bell's theorem~\cite{Bell64} with time-bin qubits. The robustness of time-bin entanglement over fiber transmission is clearly an advantage to eliminate the locality loophole. However, the detection loophole cannot be closed without using deterministic measurements in the time-bin basis. Indeed, this is essential to overcome the 83\% detection efficiency threshold needed to achieve a loophole free vio\-la\-tion of the CHSH inequality~\cite{GM87}. To eliminate the detection loophole using the schemes proposed here, one uses the dual-rail gate of Fig.~\ref{fig:dual-rail-gate} modified such that two detectors are placed on the rails right after the coupler, as shown on Fig.~\ref{fig:measurement}. To deterministically measure in any desired basis, one can set the correct values for the coupling ratio and the phase $\varphi_1$. Hence, in principle, a simultaneous closure of the detection and locality loopholes is possible using this setup. Encouragingly, we note that a noiseless photon detector with 88\% detection efficiency at 1550~nm is now available~\cite{RLMN05}. With the improvements on optical components, it is reasonable to hope that a loophole free experiment using time-bin qubits and deterministic measurements will be performed in the near future.

Before generalizing to deterministic operations on $d$-level systems, we mention that time-bin qudits can be generated with a setup similar to the one described in Fig.~\ref{fig:time-bin}, where the single photon is split into $d$ rails instead of two, all of which are time-multiplexed in one fiber using a $d$$\times$1 switch. Moreover, entangled time-bin qudits have been successfully generated using a mode-locked laser~\cite{SZG05}. 

The generalization to qudit operations is based on the following results: 1~-~Time-bin qudits can be easily converted to rail qudits by demultiplexing the bins in $d$ rails using a $1$$\times$$d$ switch and by synchronizing all the components with delay lines, 2~-~Arbitrary unitary operations on rail qudits can be implemented using 2$\times$2 couplers only~\cite{RZBB94}. 
We describe the second result in detail. Consider a single photon in a superposition of being in $d$ rails labelled 1 to $d$. Let $B_{m,n}$ be the transfer matrix of a lossless 2$\times$2 coupler mixing rails $m$ and $n$ and let $B'_{m,n}$ be the mathematical extension of $B_{m,n}$ to a $d$$\times$$d$ unitary matrix acting only on the subspace of rails $m$ and $n$. More precisely, $B'_{m,n}$ consists of the $d$$\times$$d$ identity matrix $I$ with elements $I_{mm}$, $I_{mn}$, $I_{nm}$ and $I_{nn}$ replaced by the elements of $B_{m,n}$. In Ref.~\onlinecite{RZBB94}, it is shown that any $d$-dimensional unitary transfer matrix U($d$) can be factorized in a sequence of $B'_{m,n}$ matrices. The decomposition is written as
\begin{equation} \label{eqn:decomposition}
  \mathrm{U}(d) = P \cdot \tilde{B}'_{2,1}  \ldots \tilde{B}'_{d-1,1} \cdot \tilde{B}'_{d,1} \,,
\end{equation}
where $\tilde{B}'_{d,1}$ is a sequence of $d-1$ coupler matices sequentially mixing rails $d$ and $d-1$, $d$ and $d-2$, and so on until $d$ and 1:
\begin{equation}
  \tilde{B}'_{d,1} =  B'_{d,1} \ldots B'_{d,d-2} \cdot B'_{d,d-1}\,.
\end{equation}
The decomposition is similar for $\tilde{B}'_{d-1,1}$, \ldots, $\tilde{B}'_{2,1}$. The matrix $P$ is a phase correction applied to each mode and requires $d$ phase modulators. Hence, a maximum of \mbox{$(d-1)(d-2)/2 \sim \mathcal{O}(d^2)$} couplers are necessary to implement any U($d$) in the $d$-rail encoding. As time-bin qudits can always be converted to rail qudits and vice versa, any U($d$) can be implemented in the time-bin encoding.

\begin{figure}[!h]
\includegraphics[scale=1]{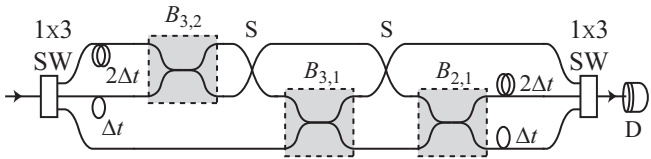}
\caption{\label{fig:QKD-qutrit} Experimental layout of the time-bin gate for QKD with qutrits. 
Lines are optical fibers, SW is a 1$\times$3 optical switch, the dashed boxes show the couplers implementing the unitary transformation on the rail qutrit and S is a rail swap to enable modes 1 and 3 (lower and upper rails) to interfere.}
\end{figure}

To exemplify our proposition, we illustrate how it can be used to implement the deterministic measurements of the quantum key distribution protocol using qutrits proposed in Ref.~\onlinecite{BPP00}. This protocol was shown to be more robust to noise than the BB84 protocol using qubits~\cite{CBKG02}. To generate a key bit, Alice chooses randomly one qutrit state among twelve that are part of four mutually unbiased basis and sends it to Bob. Let $\{\ket{a}, \ket{b}, \ket{c}\}$ be the first orthogonal basis. The second basis can be taken as
\begin{eqnarray}
  \ket{a'} &=& ( \ket{a} + \ket{b} + \ket{c} ) /\sqrt{3}\, , \\
  \ket{b'} &=& ( \ket{a} + \mathrm{e}^{2\pi i /3}\ket{b} + \mathrm{e}^{-2\pi i /3}\ket{c} ) /\sqrt{3}\, , \\
  \ket{c'} &=& ( \ket{a} + \mathrm{e}^{-2\pi i /3}\ket{b} + \mathrm{e}^{2\pi i /3}\ket{c} ) /\sqrt{3}\, .
\end{eqnarray}  
The third and fourth basis are given by the cyclic permutations of the following states, respectively:
\begin{eqnarray}
    & &( \mathrm{e}^{2\pi i /3}\ket{a} + \ket{b} + \ket{c} )/\sqrt{3}\, ,\\
    & &( \mathrm{e}^{-2\pi i /3}\ket{a} + \ket{b} + \ket{c} )/\sqrt{3}\, .
\end{eqnarray}
Upon reception, Bob measures the qutrit in one of the four randomly chosen basis. If the bases match, Alice and Bob keep the results. With time-bin qutrits, Bob can take the first basis to be the time of arrival of the photons. In this case, measurements require only one photon detector and a time frame shared with Alice. However, to measure in the second basis, Bob needs an interferometric setup to apply the unitary transformation $U\ket{\alpha'} = \ket{\alpha}$ in the time-bin encoding, where $\alpha \in \{a, b, c\}$, before measuring with his detector. The matrix $U$ is given by
\begin{equation}
    U = \frac{1}{\sqrt{3}} \left(
\begin{array}{ccc}
  1 & 1 & 1 \\
  1 & \mathrm{e}^{-2\pi i /3} & \mathrm{e}^{2\pi i /3} \\
  1 & \mathrm{e}^{2\pi i /3} & \mathrm{e}^{-2\pi i /3} 
\end{array}
\right)\, .
\end{equation}
Using equation~(\ref{eqn:decomposition}), $U$ is factorized as 
\begin{equation}
U = P \cdot B'_{2,1} \cdot B'_{3,1} \cdot B'_{3,2}\, ,
\end{equation}
where the 2$\times$2 matrices associated with $B'_{3,2}$, $B'_{3,1}$ and $B'_{2,1}$ are
\begin{eqnarray}
  B_{3,2} &=& \frac{1}{\sqrt{2}}\left(
\begin{array}{cc}
  \mathrm{e}^{i\pi /3} & 1 \\
  \mathrm{e}^{4\pi i /3} & 1 \\
\end{array}
\right)\, , \\
  B_{3,1} &=& \frac{1}{\sqrt{3}}\left(
\begin{array}{cc}
  \sqrt{2}\,\mathrm{e}^{-i\pi /3} & 1 \\
  \mathrm{e}^{-i\pi /3} & -\sqrt{2} \\
\end{array}
\right)\, , \\
  B_{2,1} &=& \frac{1}{\sqrt{2}}\left(
\begin{array}{cc}
  i & 1 \\
  -i & 1 \\
\end{array}
\right)\, .
\end{eqnarray}
The fiber optic circuit implementing $U$ is shown in Fig.~\ref{fig:QKD-qutrit}. First, a 1$\times$3 switch along with delays of $\Delta t$ and $2\Delta t$ are used to convert the incoming time-bin qutrit to a rail qutrit on which the transformation $U$ is applied using three couplers shown in dashed boxes. In the final section, the resulting qutrit is converted back to the time-bin encoding using a 3$\times$1 switch. The phase correction $P$ is not necessary in this case since detection occurs right after the gate. To measure in the third and fourth basis, it is easy to show that Bob can use the same setup supplemented with phase modulators inserted on all rails between the couplers.

In this letter, we have shown how deterministic unitary operations on single time-bin qubits encoded in photons can be implemented with current optical fiber technology. This also allows to implement deterministic measurements in any basis. Finally the generalization to deterministic operations on time-bin qudits was presented. Ultimately, the schemes are technologically limited by the loss of the switches. Since no physical reason limits the improvements of current optical switches, it is reasonable to believe that experimental quantum communication with time-bin qudits using the schemes presented here will become highly practical in the near future. Future work will focus on the experimental realization of the proposed schemes and on the application to deterministic quantum communication experiments.

\begin{acknowledgments}
Financial support by the Natural Sciences and Engineering Research Council of Canada (NSERC) and by the Canadian Institute for Photonics Innovations (CIPI) is acknowledged.
\end{acknowledgments}

\bibliography{../../felix-bibtex}

\end{document}